\documentclass[fleqn,10pt,twocolumn]{wlscirep}
\usepackage{graphicx}
\usepackage{amsmath, amssymb}
\usepackage{textcomp}
\usepackage{psfrag}
\usepackage{epsfig}
\usepackage{epsf}

\begin{document}
\title{A general model of focal adhesion orientation dynamics in response to static and cyclic stretch}

\author[1,*]{Rumi De}
\affil{Department of Physical Sciences, Indian Institute of Science Education and Research Kolkata, Mohanpur -- 741246, West Bengal, India}
\affil[*]{Correspondence and requests for materials should be addressed to R.D. (email: rumi.de@iiserkol.ac.in).}

\date{\today}

\begin{abstract}

Understanding cellular response to mechanical forces is immensely important for a plethora of biological processes. Focal adhesions are multi-molecular protein assemblies that connect the cell to the extracellular matrix and play a pivotal role in cell mechanosensing. Under time varying stretches, focal adhesions dynamically reorganize and reorient and as a result, regulate the response of cells in tissues.  Here,  I present a simple theoretical model based on, to my knowledge, a novel approach in the understanding of stretch sensitive bond association and dissociation processes together with the elasticity of the cell-substrate system to predict the growth,  stability and the orientation of focal adhesions in the presence of  static as well as cyclically varying stretches. The model agrees well with several experimental observations; most importantly, it explains the puzzling observations of parallel orientation of focal adhesions under static stretch and nearly perpendicular orientation in response to fast varying cyclic stretch.  

\end{abstract}

\maketitle

\section*{Introduction}

Mechanical forces have long been known to play an important role in determining cellular functions and behaviours \cite{SheetzReview14}. 
Living cells  actively respond to the mechanical properties of the extracellular matrix, its rigidity, also to the presence of external forces by regulating various processes such as cell adhesion, orientation, migration, differentiation, alteration in morphology and even apoptosis \cite{SheetzReview14, Geiger09, SamReview13, NicolasReview12}.
Moreover, the effect of time varying cyclic stretch is particularly striking where each cycle of contraction and relaxation leads to dynamic changes that affect a wide range of activities involving cardiovascular cells, muscle cells, stem cells, connective tissue cells to name a 
few \cite{SheetzReview14, DeReview10, Tamada04, Waterman13}.
It is not yet well understood how the stretch induces reorganization of the cellular cytoskeleton, assembly and disassembly of focal adhesions, or how it alters the gene expression, and affects the whole tissue adaptation in general.
A deeper insight into the mechanical stretch induced responses is thus envisaged to have wide impact in many cellular processes, wound healing, tissue engineering, and also regenerative medicine \cite{SheetzReview14}.

Recent researches have established that focal adhesions (FAs) 
play a crucial role in cell mechanotransduction \cite{SheetzReview14, Geiger09, SamReview13, NicolasReview12}.
FAs are micron-sized complex multimolecular protein assemblies linked on one side to the extracellular matrix via membrane-bound receptors and on other side to the actin stress fibers in the cell cytoskeleton.
Experiments show that in response to substrate stretch, focal adhesions reorganize and reorient and as a result, regulate the response of the cell \cite{Geiger09, Waterman13, Goldyn09, Huang2012, Yoshigi05, Carisey13, Greiner13}.
It is also found that external force strongly affects the growth and the stability of focal adhesion contacts \cite{SheetzReview14, Geiger09,  SamReview13, NicolasReview12}.
FAs grow in the direction of tensile stretch \cite{Balaban01} and the stability increases with the increase in applied stretch magnitude upto an optimal value, however,  eventually become unstable at higher stretches \cite{Marshall03, Vogel08}.
Moreover, another puzzling experimental observation is that focal adhesions 
respond differently to static stretch compared to rapidly varying cyclic stretch. Subjected to static or quasi-static stretch, FAs tend to orient along the stretch direction \cite{Waterman13, Liu2014, Collinsworth2000, Eastwood98}, whereas, under fast varying stretch, FAs opt to orient away from the stretch direction; for high frequency cyclic stretch, FAs align nearly perpendicular to the applied stretch direction \cite{Goldyn09, Huang2012, Yoshigi05, Carisey13, Greiner13, Kaunas09, Jungbauer08, Wang01}.

There are many theoretical studies that have contributed significantly to the understanding of how cells actively respond to the mechanical forces and regulate force transmission \cite{SamReview13, NicolasReview12}.
Quite a few studies have also been carried out to understand the stability and growth of focal adhesions in response to forces.
In a seminal work by Bell, the stability of adhesion cluster, modelled as a collection of molecular bonds, was first addressed using kinetic theory of chemical reactions. In Bell's model the rupture rate of ligand-receptor bonds was proposed to increase exponentially with the mechanical force \cite{Bell78}.
Later, the bond dissociation pathways have been studied in
the framework of Kramers theory as thermally assisted escape
over  an energy barrier under applied forces \cite{Evans97, Evans07}. 
Also, the stochasticity in  bond breaking and binding processes has been incorporated through the 
one-step master equation  to investigate the stability of focal adhesions under constant loading \cite{Schwarz04}.
However, relatively few studies have been performed on the orientational response of focal adhesions to substrate stretching. 
Moreover, existing theories, which have provided many insights into the cellular orientation problem, including our earlier works, mostly dealt with the orientation of the whole cell modelled as a contractile force dipole in a coarse-grained picture \cite{DeNphys07, DePRE08, DeBiophysJ08}; or studied the reorientation of 2D cells emphasizing on passively stored elastic energy \cite{Livne2014}; else described the formation and realignment of stress fibers in response to cyclic stretch \cite{Kaunas09, Wei08}.
Recently, Qian {\it et. al.} constructed  rate equations of the density of stress fibers and adhesive receptor-ligand bonds to describe the dynamics of cell realignment in response to cyclic stretch \cite{Qian13}. It has been hypothesized that cells tend to orient in the direction where the formation of stress fibers is energetically most favourable.
Moreover, the frequency dependent force generated within the stress fibers 
solely obtained from the structural considerations of the filament  and does not depend on the description of assembly or disassembly of focal adhesions bonds. 
However, recent experiments 
have shown that orientation specific activation of stretch sensitive proteins in FAs controls the orientation specific responses of FAs growth and disassembly \cite{Waterman13}.
Also, other studies on FAs attempted till now, 
have ignored the parallel orientation of FAs and focused only on the perpendicular orientation under cyclic stretching \cite{Chen12, Zhong11, Kong08}. 
Moreover, these studies predict that the orientational response of FAs remains unaffected under low frequency or quasi-static stretch \cite{Chen12, Zhong11, Kong08}.
Therefore, though the previous studies have provided many insights into the orientation specific response, however, so far the focus mainly remained on the dynamics of the cell, or the stress fibers, or the perpendicular orientation of focal adhesions.
Thus, a single theory which explains the parallel orientation of FAs towards 
the static stretch direction as well as the perpendicular orientation under 
fast varying cyclic stretch, remains elusive.

In this paper, a theoretical model is presented to study the orientation-specific response of focal adhesions in the presence of static as well as time varying stretches within an unified framework. 
 The crux of this model lies on a novel approach in the understanding of stretch sensitive bond association and dissociation processes of focal adhesions assembly which
play a crucial role in determining the orientational response of FAs. 
It also takes into account the elasticity of the cell-matrix system and the stochastic behaviour of ligand-receptor bonds breaking and binding processes of focal adhesions assembly. In particular, the force sensitive catch-bond behaviour and also the time-dependent binding rates under substrate stretching have been incorporated, that allow  to capture 
the experimentally observed puzzle of the parallel alignment of focal adhesions in response to static applied stretch, whereas, nearly perpendicular alignment under fast varying cyclic stretch \cite{Waterman13, Liu2014, Collinsworth2000, Eastwood98, Goldyn09, Huang2012, Yoshigi05, Carisey13, Greiner13, Kaunas09, Jungbauer08, Wang01}. 
In addition, this theory predicts several other experimental observations such as the growth and the stability of focal adhesions in the direction of tensile stretch \cite{Balaban01, Marshall03, Vogel08} and also the existence of threshold frequency and magnitude of the applied stretch that triggers reorganization of focal adhesions \cite{Kaunas09, Jungbauer08, Dartsch86}. Moreover, it also elucidates the experimental observations where orientational responses have been found to vary across cell types as a function of frequency of the substrate stretch \cite{Kaunas09, Jungbauer08, Wang01}. \\

\section*{Results}

\subsection*{Theoretical model of focal adhesions}

\begin{figure}[!h]
\includegraphics[width=8cm]{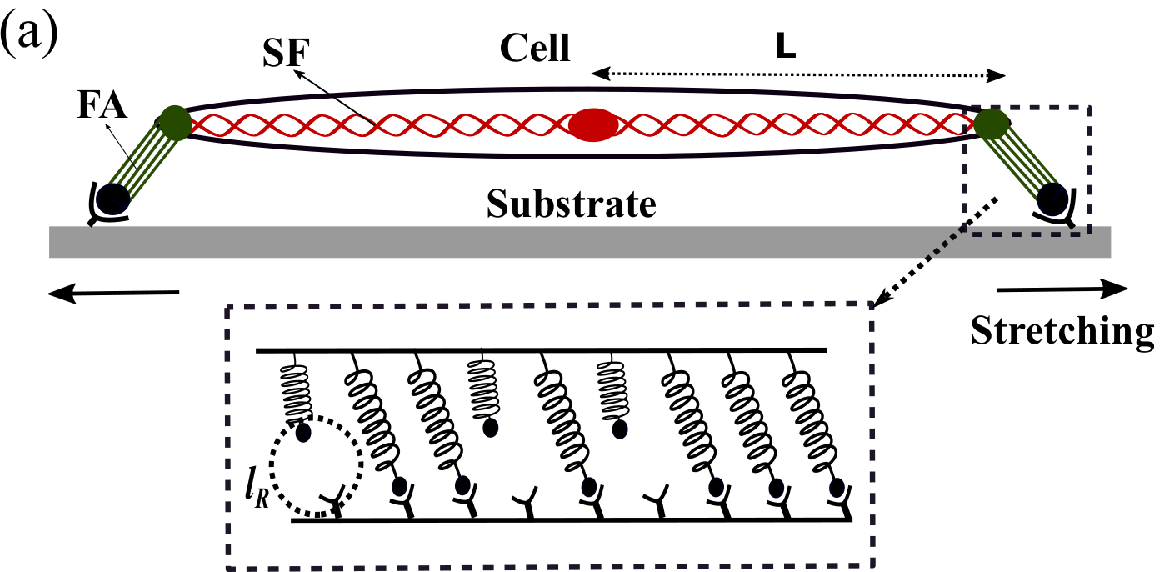}
\vspace*{5mm}
\center{
\includegraphics[width=5.5cm]{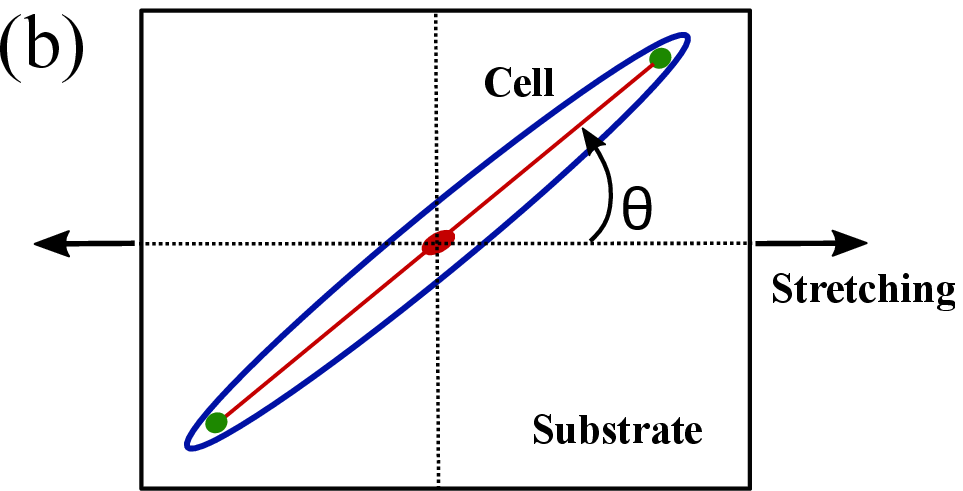}
}
\caption{(a) Schematic view of the cell and the stress fiber (SF) adhering through two focal adhesions (FA) under substrate stretching.  The inset (dotted box) illustrates the ligand-receptor bonds modelled as Hookian springs.
(b) Illustration of instantaneous orientation of the cell at an angle $\theta$ relative to applied stretching direction.
}
\label{cell_ECM_bond}
\end{figure}

\noindent
The model takes the cue from the fact that FAs are clusters of ligand-receptor molecular bonds that provide the physical connection between the cell and the extracellular matrix. Figure \ref{cell_ECM_bond}a shows a schematic representation of a cell adhering to a substrate through two focal adhesions.  
In a minimal model, it could be thought of an actin stress fiber (SF) adhered 
via two FAs. Figure \ref{cell_ECM_bond}b illustrates the instantaneous position of the cell, and hence of the focal adhesion, oriented  at an angle $\theta$ relative to the applied stretch direction. Each focal adhesion site consists of uniformly distributed ligand-receptor bonds connected in parallel to the substrate. These bonds are considered as Hookian elastic springs of stiffness, $k_{\rm b}$. 
Moreover, in this model, for simplicity, the elasticity of the cell or the stress fiber is represented by a  spring of rigidity $k_{\rm c}$ (is referred as cellular spring).
The ligand-receptor bonds are considered to be either in closed or in open state. 
Due to substrate stretching, the closed bonds, attached at one side to the substrate, get elongated and thus, experience an additional tension  (as illustrated 
in Fig. \ref{cell_ECM_bond}a); since the other side of the bonds is connected to the stress fiber, the cellular spring also gets stretched.
Thus, if the bond displacement along the stretch direction is denoted by $u_{\rm b}$ and  the cell spring displacement as $u_{\rm c}$, then the geometric constraint relation of elastic displacements is given by, $u_{\rm b}  +u_{\rm c}=L\epsilon$.
Here, $\epsilon$ is the strain magnitude and $L$ is the distance of the FA site from the cell center as shown in Fig. \ref{cell_ECM_bond}a (also, see the supplementary Fig. 1). 
Moreover, considering the force balance condition, $k_{\rm c} u_{\rm c}=\sum_{n} k_{\rm b} u_{\rm b}$;
the total force summing over all closed bonds must be balanced by the tension in the stretched cellular spring.
The above elasticity modelling enables us to determine the single bond force, 
 $f_{\rm b}$, calculated as,
$f_{\rm b}=k_{\rm b} u_{\rm b} = L\epsilon k_{\rm b}k_{\rm c}/(k_{\rm c}+ n k_{\rm b})$; $n$ denotes the number of closed bonds at any instant.

Moreover, the dynamics of focal adhesions is subjected to fluctuations in the surrounding micro-environment, thus, bonds can also undergo stochastic breaking or rebinding.
To study the time evolution of the focal adhesion cluster, a master equation has been written by coupling the elasticity of the cell-matrix system with the statistical behaviour of bond association and dissociation processes \cite{SamReview13, Schwarz04, VanKampenBook}, 
\begin{eqnarray}                                   
\frac{dP_n}{dt}=K_{\rm on}P_{n-1}+K_{\rm off}P_{n+1}-(K_{\rm on}+ K_{\rm off}) P_n, 
\label{master_eqn}
\end{eqnarray}
where $P_n(t)$ is the probability that $n$ bonds are formed at time $t$. The first two terms in the right hand side represent the gain term, i.e., the tendency for the number of bonds in state $n$ to increase due to the formation of new bond in state $(n-1)$ and the dissociation of bond in state $(n+1)$, respectively. The last term represents the loss of bonds in state $n$, 
whereas $K_{\rm on}$ and $K_{\rm off}$ denote the total association and total dissociation rate of bonds at the respective state, $n$, at any instant of time $t$.
This is further to note here that for mathematical simplicity, it has been considered that all bonds in the adhesion cluster experience the same elastic force or deformation.
However, since the rates  are subjected to 
fluctuations during stochastic simulation of the model, thus, it also takes care of the non-uniformity that may arise in bond forces across the adhesion cluster.

Moreover, during the time evolution, the bond reaction rate strongly depends on the instantaneous bond configuration. 
Recent single molecule experiments have revealed that the external force increases the lifetime of many FAs molecules \cite{Kong09, RocaCusachs09}.  Tensile force, upto an optimal value, is found to strengthen and reinforce the molecular bonds; and bonds' lifetime decreases with further increase in force \cite{Marshall03}. These type of force strengthening bonds are called catch bonds and are believed to play a crucial role in stabilizing the focal adhesion cluster \cite{Vogel08}.
Motivated by the experimental findings, in this model, it is assumed that the FA cluster consists of catch bonds and thus, the dissociation rate, $k_{\rm off}$, of  the closed bond is proposed to demonstrate the catch behavior as \cite{Pereverzev05,  Vogel08},                                                     
\begin{equation}
k_{\rm off}=k_{\rm slip}e^{f_{\rm b}/f_{\rm 0}}+k_{\rm catch}e^{-f_{\rm b}/f_{\rm 0}};
\label{catchoffrate}
\end{equation}
where $k_{\rm slip}$  and $k_{\rm catch}$ denote the rate constants for dissociation of the ligand-receptor pair via a slip pathway promoted by the force and a catch pathway opposed by the force respectively \cite{Pereverzev05};  these rate constants depend on the type of the adhesion molecules. 
The total dissociation rate, $K_{\rm off}$,  is thus,  $K_{\rm off}=\sum_n k_{\rm off}$, where $n$ is the number of closed bonds at any given instant
and $f_{\rm 0}$ denotes a molecular force scale, typically of the order of piconewton.

On the other hand, it has generally been considered  that the association or rebinding rate, $k_{\rm on}$, increases with the number of available unbound bonds at any instant; thus,
$k_{\rm on}=\gamma (N-n)$, 
where $N$ is the total number of binding sites, $n$ is the number of closed bond, and $\gamma$ is the binding rate constant \cite{Schwarz04}.
However, in this model, motivated by the experiments \cite{Robert09, Robert11}, it is considered that 
during the rebinding process, to facilitate the ligand-receptor bond formation, the pair needs to be in close proximity within a reaction radius, $l_{\rm R}$, and for a characteristic contact time, $t_{\rm R}$.
 Thus, the ligand-receptor bond has an intrinsic reaction rate,  $v_{\rm 0}=l_{\rm R}/t_{\rm R}$, for binding to occur.
 Therefore, in the presence of a cyclically varying substrate stretch, as the substrate moves back and forth, the ligand attached to the substrate also moves back and forth from the cell receptor and hence the association or rebinding rate strongly depends on the time variation of the applied stretch. 
Thus, under a cyclic strain, $\epsilon(t)=\epsilon_{\rm 0}(1-\cos \omega t)$, 
where $\epsilon_{\rm 0}$ is the average magnitude and  $\omega$ is the frequency of the applied stretch;
the ligand which was initially at a distance $l_{\rm R}$ from the cell receptor would now undergo a cyclically varying displacement, $u(t)=l_{\rm R} \epsilon(t)$; and
hence, continuously moves away back and forth from the cell receptor. 
Therefore, the displacement rate, $|\dot u(t)|=l_{\rm R} |\dot{\epsilon}(t)|$ has to be much smaller compared to the intrinsic binding rate, $v_{\rm 0}$, 
so that the ligand receptor pair gets sufficient time to rebind. 
If the absolute magnitude of the displacement rate, $v_{\omega}=l_{\rm R}\epsilon_{\rm 0} \omega <<v_{\rm 0}$,
the ligand-receptor pair is in contact for long enough time so that the reaction can take place and binding can easily occur.
However, as the displacement rate increases with increase in stretching magnitude or frequency of the external stretch, the ligand-receptor pair gets to spend less and less contact time as the ligand moves away rapidly from the cell receptor
and therefore the probability of bond formation  decreases. 
To incorporate these effects, in this model,  the association rate is proposed to be rate-dependent and described by
\begin{equation}
k_{\rm on}= \gamma (N-n) e^{-v_{\omega}^2/v_{\rm 0}^2}.
\label{konrate}
\end{equation}

Also, this is to note here that even though, $l_{\rm R}$, does not explicitly appeared in the expression of $k_{\rm on}$,
however, the effect of the reaction radius could be seen through its dependence on the strain magnitude, $\epsilon_{\rm 0}$.
This dependence  solely appears due to the consideration of the reaction radius. Therefore, in this model, the effective time scale turns out to be, $t_{\omega}=1/(\epsilon_{\rm 0} \omega)$. 
Thus, the competition between the two time scales, the time variation of the substrate displacement, $t_{\omega}$, and  the intrinsic binding time scale, $t_{\rm R}$,  determines the probability of bond formation.

Now, in the presence of a static stretch, the association rate  
turns out to be, $k_{\rm on}= \gamma (N-n)$, as $\omega=0$. 
Under a static stretch, the distance between the ligand-receptor pair increases with the external stretch; however,  since the pair
spends sufficiently long time in close contact, thus, there is always chances of  ligand-receptor rebinding to occur and hence the adhesion bonds could be formed. In case of a static stretch, the dissociation rate, $k_{\rm off}$,
plays a major role in determining the stability of the adhesion cluster. 
Due to the catch bond behaviour of the dissociation rate, 
FAs get strengthen with tensile stretch, 
however, above a threshold stretch value, the dissociation rate start increasing and eventually wins over  the association rate and hence the cluster becomes unstable and thus, disassemble.

\subsection*{Numerical simulation method}
All parameters of the model have been rewritten in dimensionless units. The scaled time is defined as $\tau=k_{\rm 0} t$, 
where $k_{\rm 0}$ is the spontaneous dissociation rate in the absence of any external force. Similarly, all other rates have been scaled, such as, $K_{\rm s}=k_{\rm slip}/k_{\rm 0}$, $K_{\rm c}=k_{\rm catch}/k_{\rm 0}$, $\Gamma=\gamma/k_{\rm 0}$, $T_{\rm R}=k_{\rm 0}t_{\rm R}$ and the scaled frequency, $\omega=\omega_{\rm 0}/k_{\rm 0}$. The normalised bond force is defined as $F_{\rm b}=f_{\rm b}/f_{\rm 0}$. 
Moreover, all displacements have been scaled by the characteristic length, $l_{\rm R}$, such that $U_{\rm b}=u_{\rm b}/l_{\rm R}$, $U_{\rm c}=u_{\rm c}/l_{\rm R}$, and $L^{\rm s}=L/l_{\rm R}$.

The master equation has been numerically solved to investigate the time evolution, growth, stability and the orientation of adhesion clusters in response to static as well as time varying cyclic stretch.
Monte-Carlo method has been used based on Gillespie's algorithm \cite{Gillespie}. 
In simulations, the FA cluster consisting of $N$ binding sites, starts from an initial state with all bonds at closed state and proceeds until all bonds become open. 
Thus, the stochastic trajectories are simulated which share many similarities with experimental realizations.  Averaging has been done over many such trajectories to extract useful statistical information.
In case of stable clusters, each simulation has run for at least one million events. 
The overall statistics have been accumulated from 500 such simulation results.
In this model, following the existing literature, stability and growth of the adhesion cluster is represented by the mean number of closed bonds under different conditions such as varying stretch, frequency etc.
Dynamics have been studied for a wide range of parameter values. Here, the results presented for a cluster of size N=200. The length $L$ is taken as $20 \mu m$ and $l_{\rm R} \sim 1 nm$. The ratio of single bond stiffness to cellular spring is taken as $k_{\rm b}/k_{\rm c}=5$. 
The other scaled parameter values are kept at $\Gamma=2$, $K_{\rm s}=0.10$, and 
$K_{\rm c}=120$ (following \cite{Pereverzev05}).
The magnitude and frequency of the applied stretch and $T_{\rm R}$ remain variable parameters in the model.

\begin{figure}[!t]
\center{
\includegraphics[width=7.0cm]{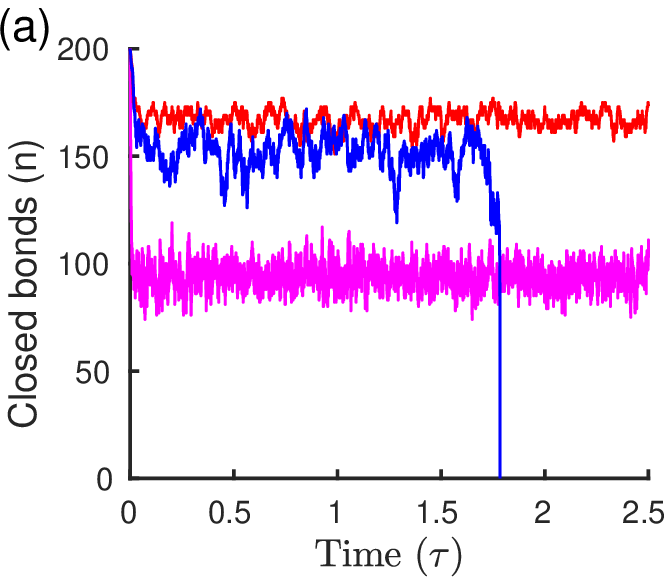}
\vspace{7mm}
\includegraphics[width=7.0cm]{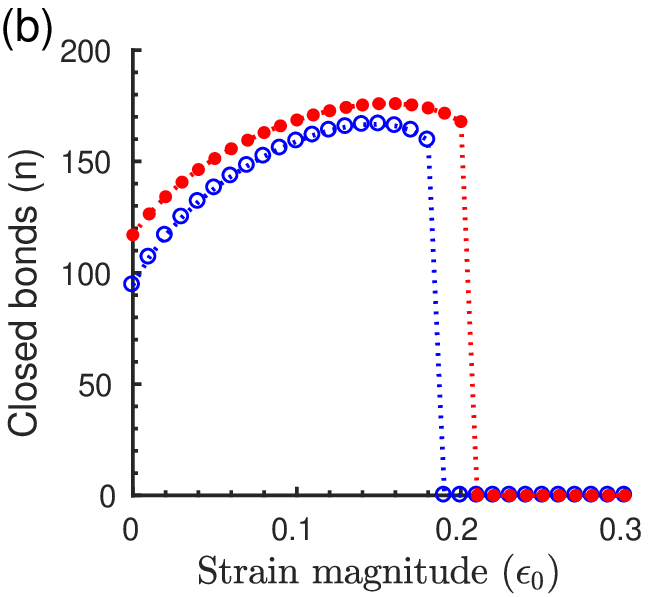}
}
\caption{Simulation results for static stretch. (a) Time evolution of mean number of closed bonds for three different strain magnitudes: $1\%$ (pink), $10\%$ (red) (stable cluster), and $20\%$ (blue: unstable).
(b) Mean number of closed bonds as a function of strain magnitude for $\Gamma=1$ (open blue circles) and $\Gamma=2$ (solid red dots). Cluster size increases with increase in $\Gamma$ value. 
}
\label{fig-staticstretch}
\end{figure}

\subsection*{Orientational response in the presence of static stretch}
Figure \ref{fig-staticstretch}a shows the typical simulation trajectories of instantaneous number of closed bonds for three different magnitudes of static strain. 
It is observed that the number of closed bonds stochastically varies around a mean value depending on the applied stretch magnitude and the cluster  eventually becomes unstable above a threshold stretch. 
The effective strain magnitude along the FA cluster oriented at an angle $\theta$ is $\epsilon_{\rm a} =\epsilon_{\rm 0} \cos^2 \theta$ (as in Fig. \ref{cell_ECM_bond}b).
The stability and the growth of the adhesion cluster (represented by the mean number of closed bonds) along the direction of the applied static stretch (i.e., $\theta=0$) have been studied as a function of strain magnitude 
as shown in Fig. \ref{fig-staticstretch}b. 
It is found, as observed in experiments, the adhesion cluster grows with increasing strain; reaches a maximum under an optimal strain value, however,  a further increase in stretch eventually results in disassembly of the cluster. This could be attributed to the catch behaviour of FA molecules.
Since the catch bonds in FAs get strengthen with tensile stretch, the dissociation rate decreases and 
that, in turn,
promotes the growth and stability of FAs. The cluster becomes most stable for an optimal stretch at which the bond dissociation rate is minimum.
Thus, in the presence of a static stretch, below the threshold magnitude, parallel  to the stretch direction becomes the most stable direction for FAs growth; whereas perpendicular direction is the least stable 
(as $\epsilon_{\rm a} =0$) associated with high dissociation rate of bonds. Therefore, FAs tend to align along the parallel direction of the applied stretch.
The findings, thus, explain the recent experimental observation where FAs initially oriented at perpendicular direction, disassemble under a static stretch and reorient towards the parallel direction of higher stability; whereas FAs initially oriented parallel to the applied stretch direction remain stable \cite{Waterman13}.

\begin{figure}[!t]
\center{
\includegraphics[width=7.0cm]{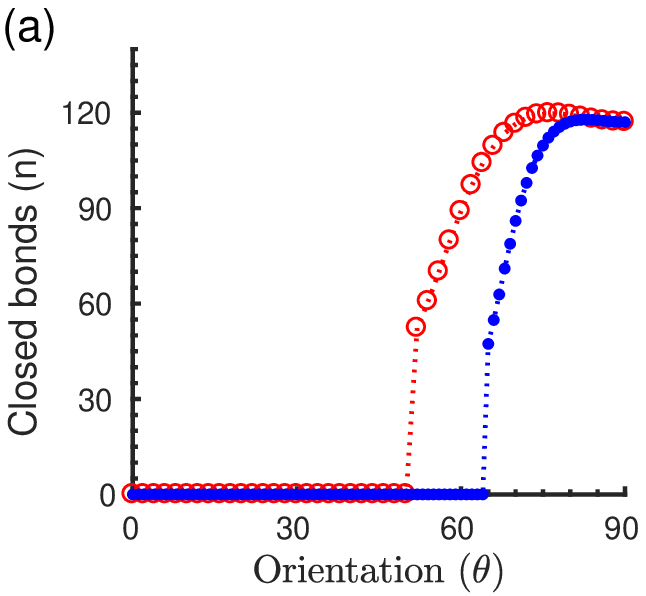}
\vspace{7mm}
\includegraphics[width=7.0cm]{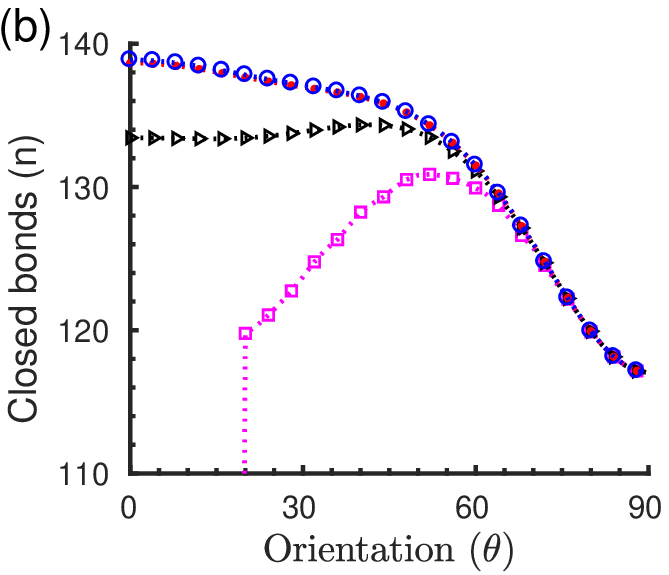}
}
\caption{Simulation results for time varying stretch. (a) Stability of the adhesion cluster as a function of orientation angle, $\theta$ (in degree), under fast varying cyclic stretch (when $T_{\omega}<T_{\rm R}$) with $\omega=10$ (solid blue dots), and $\omega=5$ (open red circles). 
(b) The plot shows the cluster stability with decrease in stretching frequency for
$\omega=1$ (open magenta square), $\omega=0.5$ (open black triangle), 
$\omega=0.1$ (solid red dots), and $\omega=0.01$ (open blue circles),
keeping $T_{\rm R}=10$ and for $10\%$ stretch.
}
\label{average-bonds-theta}
\end{figure}

\subsection*{Orientational response in the presence of dynamic stretch}
However, in the presence of a high frequency cyclic stretch,  
orientational response of focal adhesions is found to be quite different \cite{Goldyn09, Huang2012, Yoshigi05, Carisey13, Greiner13, Kaunas09, Jungbauer08, Wang01}.
Under a cyclically varying stretch, $\epsilon(\tau)=\epsilon_{\rm 0}(1-\cos \omega \tau)$, the competition between the two time scales, $T_{\omega}=1/\epsilon_{\rm 0} \omega$, 
 the time variation of the substrate displacement compared to $T_{\rm R}$, the intrinsic binding time scale of the ligand-receptor pair determines the formation and stability of FAs cluster.
For high frequency stretch, if $T_{\omega}<<T_{\rm R}$, the ligand-receptor pairs do not get sufficient contact time to rebind as the ligand on the substrate moves away rapidly from the cell receptor. 
Therefore, the chances of new bond formation decreases with increasing frequency and the FA cluster becomes unstable along the stretch direction. 
However, if FA orients, away from the stretch direction, at an angle $\theta$, then the effective 
strain magnitude, $\epsilon_{\rm a}=\epsilon_{\rm 0} \cos^{\rm 2} \theta$, acting along the FA decreases and hence
the contact time between the ligand-receptor pairs, $T_{\omega}$ increases with $T_{\omega} \propto 1/\epsilon_{\rm a}$.
Therefore, as the cluster orients away from the stretch direction, it becomes more and more stable with increase in probability of bond formation.
At perpendicular direction, which is the zero strain direction, the binding of ligand-receptor pair is no longer affected by the fast varying stretch and thus, FAs become stable.
Figure \ref{average-bonds-theta}a shows the stability of the adhesion cluster as a function of orientation angle, $\theta$, under a high frequency (for $T_{\omega}<<T_{\rm R}$) cyclic stretch.
As seen from the figure, cluster grows in size as it orients near  perpendicular direction. There exists a distribution of orientation angles towards the perpendicular direction as found in experiments.
However, as the stretching frequency decreases, ligand-receptor pairs get longer contact time and the ratio of the two time scales, $T_{\omega}$ and $T_{\rm R}$, determines the most stable orientation angle. When $T_{\omega}>>T_{\rm R}$, as the substrate moves slowly,  ligand-receptor pairs get sufficient time to form new bonds and therefore, FAs, due to strengthening of the catch bonds, tend to align along the maximal stretch direction ($\epsilon_{\rm a} =\epsilon_{\rm 0}$ for $\theta=0$), i.e., parallel to the applied stretch, as shown in Fig. \ref{average-bonds-theta}b.

\begin{figure}[!t]
\includegraphics[width=7.8cm]{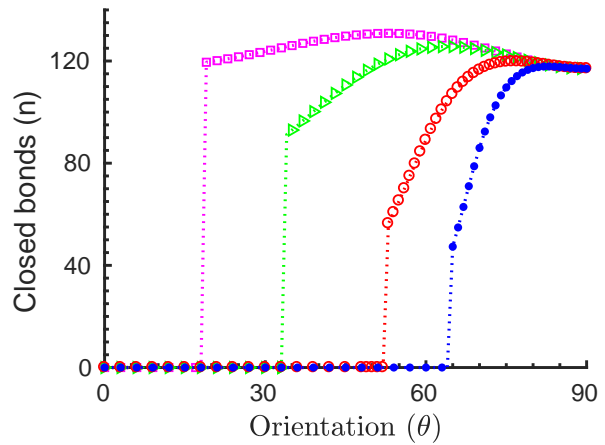}
\caption{Effect of intrinsic binding time scale, $T_{\rm R}$, on the cluster orientation for different values of $T_{\rm R}= 10$ (solid blue dots), $5$ (open red circles), $2$ (open green triangle), and $1$ (open magenta square); for $10\%$ stretch and frequency, $\omega=10$.
}
\label{average-bonds-intrinsic-bindingtime}
\end{figure}

\subsection*{Effect of intrinsic binding time scale on focal adhesion orientation}
Moreover, this theory also elucidates the experimental observations where orientational responses 
have been found to vary from cell type to cell type under time-dependent cyclic stretches. It is observed that
some cell types prefer near perpendicular alignment, whereas some orient at different angles, or some do not exhibit considerable reorientation below a certain frequency \cite{Kaunas09, Jungbauer08, Wang01}. 
In this model, it could be attributed to different intrinsic time scales, $T_{\rm R}$,  characteristic to different cell types. 
Figure \ref{average-bonds-intrinsic-bindingtime} shows the orientational stability of the adhesion cluster for different  $T_{\rm R}$ values while keeping the magnitude and the frequency of the cyclic stretch constant. 
As seen from the figure, with decrease in the characteristic rebinding time, $T_{\rm R}$,  the cluster becomes  stable at a wide range of orientation angles. As $T_{\rm R}$ decreases, since the binding occurs at a faster rate, ligand-receptor pairs could increasingly cope up with the varying substrate stretch; and thus, the cluster becomes stable.
Therefore, focal adhesions and thus, cells with fast binding rates do not show significant reorientation.

\begin{figure}[!t]
\includegraphics[width=8.0cm]{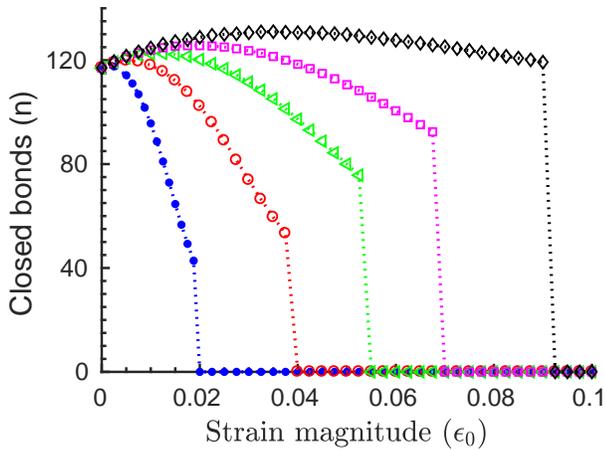}
\caption{The stability of the adhesion cluster as a function of strain magnitude, $\epsilon_{\rm 0}$, with varying frequency as,  $\omega=10$ (solid blue dots), $5$ (open red circles), $3$ (open green triangle), $2$ (open magenta square), and $1$ (open black diamond); keeping $T_{\rm R}=10$.}
\label{average-bonds-threshhold_stretch}
\end{figure}

\subsection*{Existence of threshold stretch magnitude}
Moreover, as found in experiments, this theory also captures the existence of a threshold stretch magnitude above which the orientational response becomes prominent \cite{Kaunas09, Jungbauer08, Dartsch86}.
Figure \ref{average-bonds-threshhold_stretch} shows the stability of the adhesion cluster, oriented parallel to the applied stretch direction ($\theta=0$),  as a function of strain magnitude and for different stretching frequencies. 
While keeping the frequency constant, if the magnitude of the applied stretch decreases, 
since ($T_{\omega}\propto 1/\epsilon_{\rm a}$), this leads to increase in the contact time
between the ligand-receptor pair. As a result, for smaller stretch magnitudes, below a threshold value, when $T_{\omega}>T_{\rm R}$, since the probability of bond formation increases with increasing contact time, hence, the adhesion cluster becomes stable.
However,  above the threshold stretch, as $T_{\omega}<T_{\rm R}$, ligand-receptor pairs do not get sufficient contact time to rebind with the fast varying substrate; thus, the cluster becomes unstable and so orients away from the stretch direction.  Therefore, orientational response of FAs becomes prominent above a threshold stretch value.
Moreover, with increase in stretching frequency, 
as shown in Fig. \ref{average-bonds-threshhold_stretch},  since $T_{\omega}$ decreases with increasing $\omega$, ($T_{\omega}\propto 1/\omega$), the magnitude of the threshold stretch shifts to a  lower value.

\section*{Discussion}

In summary, the theoretical model presented,  by incorporating the catch-bond behaviour of focal adhesion assembly and the  time dependent binding rates under substrate stretching, agrees well with several experimental observations. Apart from capturing the force sensitive stability of focal adhesion clusters under tensile stretch, this model in an unified framework, also unravels the puzzling observation of orientation of FAs  along the parallel direction in response to static and quasi-static stretch, as well as the perpendicular orientation under fast varying cyclic stretch. 
Moreover, it explains the variation in alignment angles in different cell types and also predicts the existence of threshold stretch as observed in experiments. As discussed,
the competition between the time scales involved in the process, namely, the time variation of the substrate displacement and the intrinsic binding time  of the ligand-receptor pairs determines the stability of focal adhesions under time varying stretches.
Importantly, it is shown that the sole consideration of the stretch  dependent association and dissociation processes of adhesion clusters could successfully predict  the orientational response of focal adhesions.
Indeed, it has been shown in recent experiments 
that the orientation specific activation of stretch sensitive proteins in FAs play a crucial role in determining the orientation specific focal adhesions growth and disassembly \cite{Waterman13}.
This is also to note that the complex mechanisms of viscoelasticity of stress fibers and actin-myosin contraction may affect the overall cellular mechanosensing processes \cite {Deguchi06, Kumar06}. 
Thus, adding these effects one would get an inclusive picture, however, it is outside  the present theory. 
While this model correctly predicts the major experimental features on the orientation of focal adhesions reported so far, the finer predictions could be further tested by suitable experiments.

\section*{Acknowledgements}
The author thanks Dr. Rangeet Bhattacharyya for insightful discussions
and also acknowledges the financial support from  Science and Engineering Research Board (SERB), Grant No. SR/FTP/PS-105/2013, Department of Science \& Technology (DST), India.

\section*{Author contributions}
R.D. developed the model, carried out the entire work, and wrote the manuscript.

\section*{Additional information}
{\bf Supplementary Information} accompanies this paper.\\

\noindent
{\bf Competing interests:} The author declares no competing interests.\\

\noindent
{\bf Data availability:} The data supporting the findings of this study are available within the paper and its supplementary information file.

\end{document}